\documentclass[aps,twocolumn,groupedaddress]{revtex4}
\usepackage[section]{placeins}        
\usepackage{float}

\usepackage{float}
\usepackage{ulem}
\usepackage{caption}
\usepackage{epsfig}
\usepackage{pdfpages}
\usepackage{amsmath,amsthm, amssymb, latexsym}
\usepackage{color}
\usepackage{graphicx}
\usepackage{epsfig}
\usepackage[outercaption]{sidecap}

\newcommand{\bee}{\begin{equation}}
\newcommand{\ene}{\end{equation}}
\newcommand{\beea}{\begin{eqnarray}}
\newcommand{\enea}{\end{eqnarray}}

\begin{document}
\title{Localized absorption of laser energy in X-mode configuration of magnetized plasma}

\author{Ayushi Vashistha$^{1,2}$}
\thanks{ayushivashistha@gmail.com}
\thanks{Current Affiliation: Applied Materials Inc.}
\author{Devshree Mandal$^{1,2}$}
\thanks{Current Affiliation: Applied Materials Inc.}
\author{Srimanta Maity$^{3}$}
\author{Amita Das$^{3}$}
\thanks{amitadas3@yahoo.com}

\affiliation{$^1$ Institute for Plasma Research, HBNI, Bhat, Gandhinagar - 382428, India } 
\affiliation{$^2$ {Homi Bhabha National Institute, Mumbai, 400094 } }

\affiliation{$^3$ Physics Department, Indian Institute of Technology Delhi,  Hauz Khas, New Delhi - 110016, India }

\begin{abstract}
  The heating of ions via lower hybrid waves has been observed in several astrophysical as well as laboratory plasmas. We have conducted Particle-In-Cell simulations to demonstrate  absorption of the incident laser pulse at a chosen localized point in the target by manipulating the plasma density profile. We show that a part of the incident laser propagates inside plasma target, when its frequency lies below the lower hybrid resonance frequency. Thereafter, as it experiences a negative density gradient, it approaches the resonance point where its  group velocity approaches zero. This is where the electromagnetic energy prominently gets converted into electrostatic and eventually   into kinetic energy of ions. Thus by  tailoring the plasma  density profile one can  have the  absorption of incident electromagnetic wave energy   at a designated  location inside the  plasma.  This may have  importance in various applications  where energy deposition/heating of plasma at a localized region is desirable. 

\end{abstract}

\maketitle

\section{Introduction}
 Interaction of electromagnetic (EM) waves with plasma has been of great interest for fundamental as well as applications of plasma physics  \cite{kaw_RMP, chopineau2019identification, mandal2020spontaneous}. Interaction of laser with magnetised plasma, has also gained popularity due to its relevance in several laboratory experiments as well as space plasma observations \cite{fisch1978confining, fisch1987theory, mandal2021electromagnetic, maity2021harmonic, horton2013penetration, maity2022mode,   pinsker2015whistlers,goswami2021ponderomotive}. In laboratory plasma, the study of EM wave interaction with magnetised plasma can help in understanding phenomenon in toroidal tokamak devices and several linear devices.  Interaction of laser with plasma target embedded in external magnetic field is also being explored in Inertial Confinement Fusion (ICF) experiments \cite{Montgomery2015, Chang2011_prl_ICF}.  For instance, application of external magnetic field timed to coincide with the laser beams compressing the target in ICF configuration,   has   demonstrated improved energy gain \cite{Chang2011_prl_ICF, Hohenberger2012_ICF}.   Recently, interaction of laser with plasma in the presence of longitudinal as well as azimuthal magnetic field of the order of Mega Gauss (MG) has also been reported \cite{Ivanov2021_tpd}. They studied plasma expansion under the applied magnetic field and demonstrated excitation of two-plasmon decay instability in their system. There have also been a study on the shape of the laser produced plasma, expanding in an externally applied magnetic field \cite{Ivanov2017_disc_plasma}. 
 

 We now address the question of coupling of laser energy to plasmas.  The conventional laser plasma interaction physics leads to the laser energy  coupling with   the lighter electron species.  The secondary process of  electron ion interaction ( which could be either collisional and/or collective) is then responsible for transferring the electron  energy to ions.  In  our  recent publications, \cite{vashistha2020new, vashistha2021excitation} we have demonstrated using PIC simulations that a laser interacting with a magnetized plasma on the other hand can excite electrostatic lower hybrid waves.   This mechanism leads to the direct coupling of laser energy  to  ions. In these studies the laser was allowed to fall on a uniform density plasma from vacuum. The lower hybrid frequency essentially defines the boundary of the pass and the stop band of the X-mode electromagnetic wave propagation in the plasma. The laser frequency, therefore,  had to be chosen slightly lower than the  lower hybrid frequency for it to remain in  the pass band of X-mode.  It was observed that the  closer the  laser frequency lie to lower hybrid resonance frequency, the higher the reflection of the field. On the other hand if the laser frequency was chosen  much less than the resonance frequency, the entire pulse propagates as an electromagnetic wave with little or no conversion to electrostatic component. Furthermore, the energy absorption in these studies occurs in a bulk region of the plasma and not at localised point in the target.

Ion heating due to lower hybrid emissions has also been reported in the astrophysical as well as laboratory plasmas \cite{ion_heating, retterer_ion_suprauroral}.  For instance, observation of energetic ion  at the well defined central plasma density in Alcator A Tokamak \cite{schuss1981lower},  a rise in off axis ion temperature due to lower hybrid heating in HT-7 Tokamak \cite{liu2001ion}, heating of ions species near lower hybrid frequency in FM-1 spherator \cite{hawryluk1976rf} are few examples of the same.  However,  study of lower hybrid wave for  current drive purpose has gained more popularity in toroidal devices, due to its high efficiency \cite{fisch1978confining, fisch1987theory, pinsker2015whistlers}. Heating of ions in tokamaks via lower hybrid take place off the central axis \cite{wesson2011tokamaks} and hence it is undesirable candidate for heating purposes. But it can be possible to heat ions in linear devices. With the technological advancements in achievable magnetic field, it is viable to study such effects in laser-plasma interactions. In view of this, we explore the heating of ion species by laser at the lower hybrid resonance point in plasma.

We have considered a one-dimensional simulation geometry with two different plasma density profiles (fig. \ref{density_profile3}). The details on these profiles  will be discussed in the sections to follow. We consider (I) a constant plasma density and (II) a constant and then falling plasma density profile.   We show that laser propagates predominantly as an electromagnetic mode when its frequency lies below the LH resonance frequency. Further, as the propagating EM mode encounters a negative plasma density, its energy gets converted to electrostatic lower hybrid oscillations  at the resonance point. The lower hybrid oscillations at the resonance point then break to impart its energy into ion species. This way exchange of laser energy takes place directly into ion species at a specified location.  The details of this mechanism will be discussed in the sections to follow.

\section{Simulation Details}
We perform 1-D simulations with Particle-In-Cell  using OSIRIS-4.0 \cite{hemker,Fonseca2002,osiris}.
We consider a 1-dimensional simulation box with $x$ ranging from $0$ to $4000$ $c/\omega_{pe}$ with plasma boundary starting from $x=850$.  Grid and time resolutions are $dx= 0.05$ and $dt = 0.02$ respectively. We consider both electron and ion species in our system. The mass of ions species is considered $100$ times to the mass of electrons. This is to make the computation faster. We have considered two different plasma density profiles (fig. \ref{density_profile3}). We choose a plane polarised laser propagating along $\hat{x}$ with its magnetic field along $\hat{z}$ and electric field along $\hat{y}$. Additionally, we have applied an external magnetic field along $\hat{z}$ which makes it X-mode configuration study. We choose 8 particles per cell. The boundaries are absorbing for both particles and fields. The fields are normalised to $mc\omega_{pe}/e$. Time and length are normalised to $\omega_{pe}^{-1}$ and $c/\omega_{pe}$ respectively.  The details of the simulation parameters are given in table \ref{simulation_parameters}. 

	\begin{table}
		\caption{ Values of simulation parameters in normalised and corresponding standard units for $CO_2$ laser. }
		\begin{tabular}{|p{1.5cm}||p{2.5cm}||p{2.5cm}|}
			\hline
			\textcolor{red}{Parameter}& \textcolor{red}{Normalised Value}& 	\textcolor{red}{Value in standard unit}\\
			
			\hline
			\hline
			\multicolumn{3}{|c|}{\textcolor{blue}{Plasma Parameters}} \\
			\hline
			$n_o$&1 &$3 \times 10^{21}$ $ cm^{-3}$\\
			\hline
			$\omega_{pe}$&1&$3.3 \times10^{15}$Hz\\
			\hline
			$\omega_{pi}$&$0.1 \omega_{pe}$ (for M/m = 100) &$0.34 \times 10^{15}$Hz\\
			\hline
			\multicolumn{3}{|c|}{\textcolor{blue}{Laser Parameters}} \\
			\hline
			$\omega_l$&$0.06 \omega_{pe}$&$0.2 \times 10^{15}$Hz\\
			\hline
			$\lambda_{l}$& $104.7 c/ \omega_{pe}$ &$9.42\mu m$\\
			\hline
			Intensity&$a_{0} =0.5$&$3.5\times 10^{15} W/cm^2$ \\
			\hline
			

			\hline	
				\end{tabular}
					\label{simulation_parameters}
	\end{table}

\begin{figure}
	\centering
		
	\includegraphics[width=0.99\linewidth]{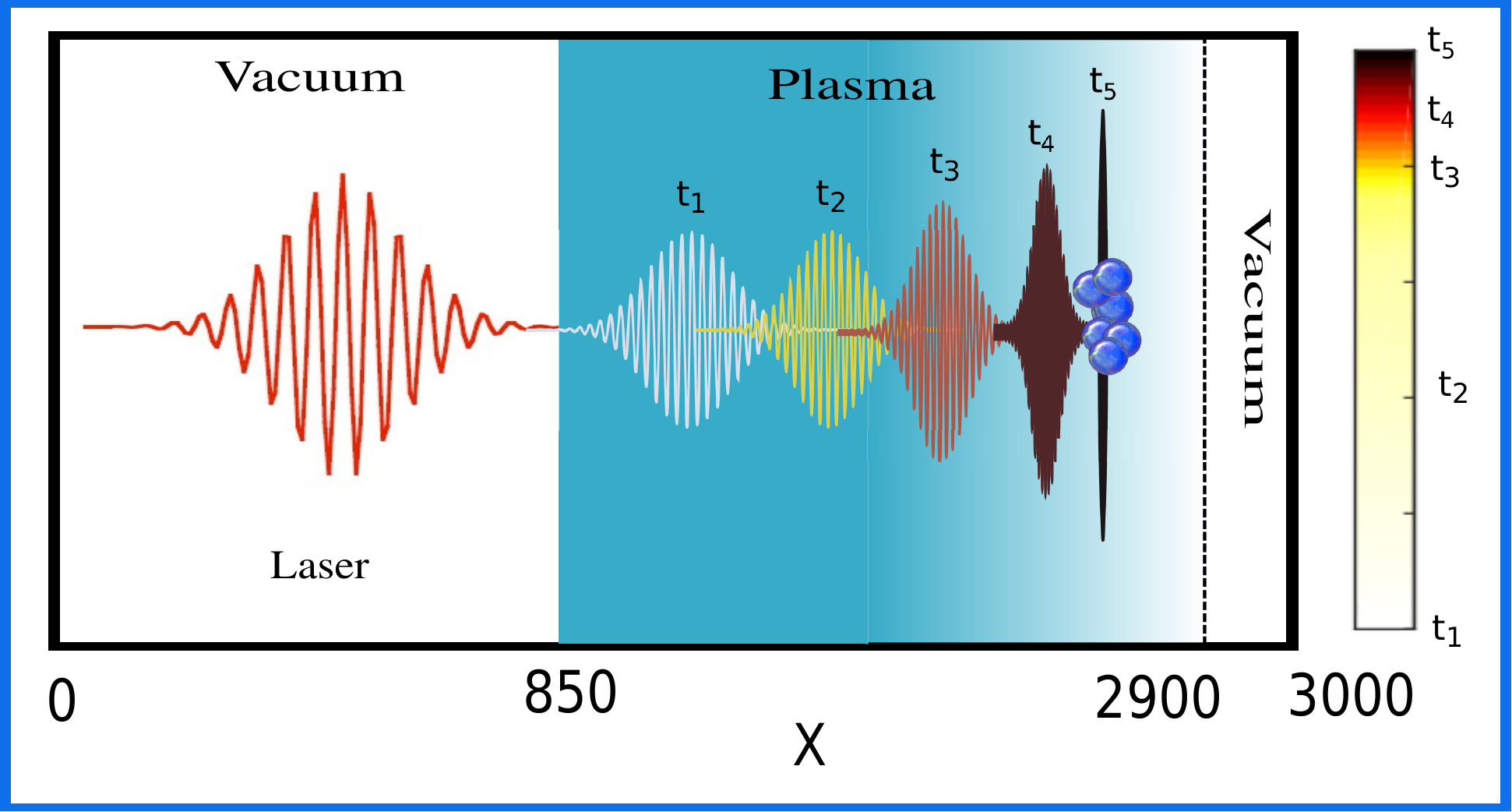}
	\caption{A pictorial representation (not to scale) of the proposed mechanism has been shown in the figure. It is shown that a  part of laser propagates inside plasma until it encounters the lower hybrid resonance point. At this point the field energy gets converted to kinetic energy of ions. }
		\label{schematic}
\end{figure}

The parameters given in table \ref{simulation_parameters} are for $CO_2$ laser. We choose external magnetic field high enough to magnetise the electrons i.e.  we maintain the condition $\omega_{ce}> \omega_{l}> \omega_{ci}$.  We choose $\omega_{ce} = 2.5 \omega_{pe}$ (in normalised units) which in real units will correspond to $46.6KT$.  The results presented in our study are however applicable to any incident electromagnetic wave frequency.  The required field parameters will change with plasma density, ion mass and incident laser frequency. The necessary condition to be maintained is that the electrons are magnetised whereas ions are unmagnetised in collisionless plasma. A schematic for the mechanism in operation is shown in fig. \ref{schematic}

\section{Results and Discussions}


\begin{figure*}
	\centering
	\includegraphics[width=1.0\linewidth]{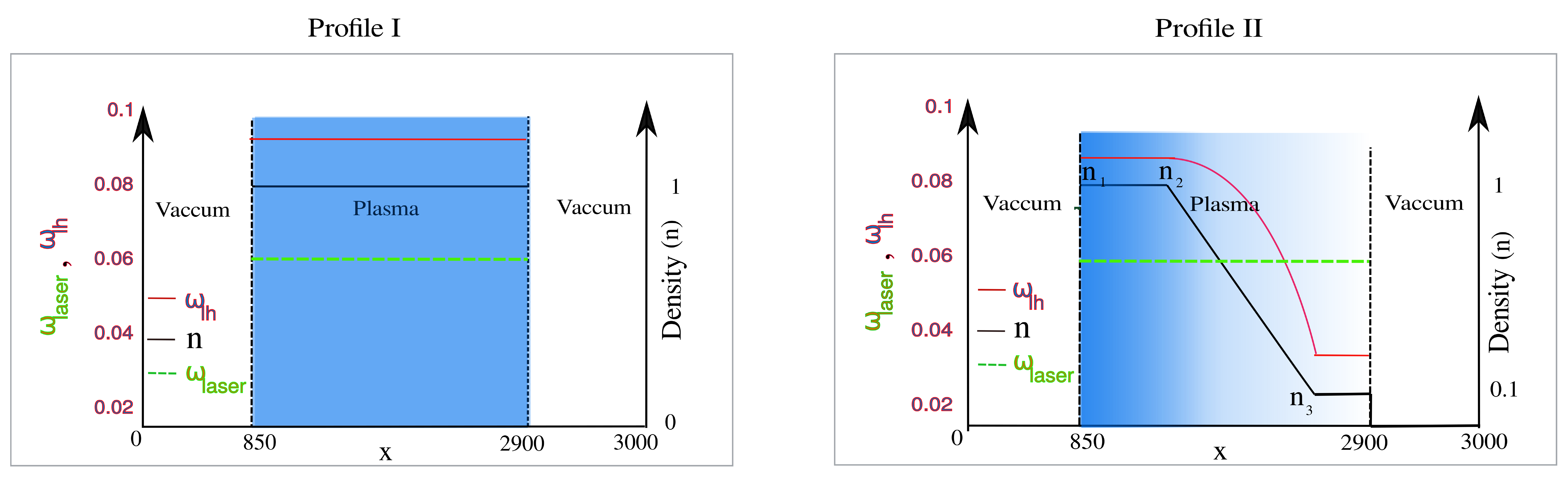}

	\caption{Schematic for different plasma density profiles considered in simulation. }
		\label{density_profile3}
\end{figure*}

\begin{figure}
	\centering
	\includegraphics[width=0.99\linewidth]{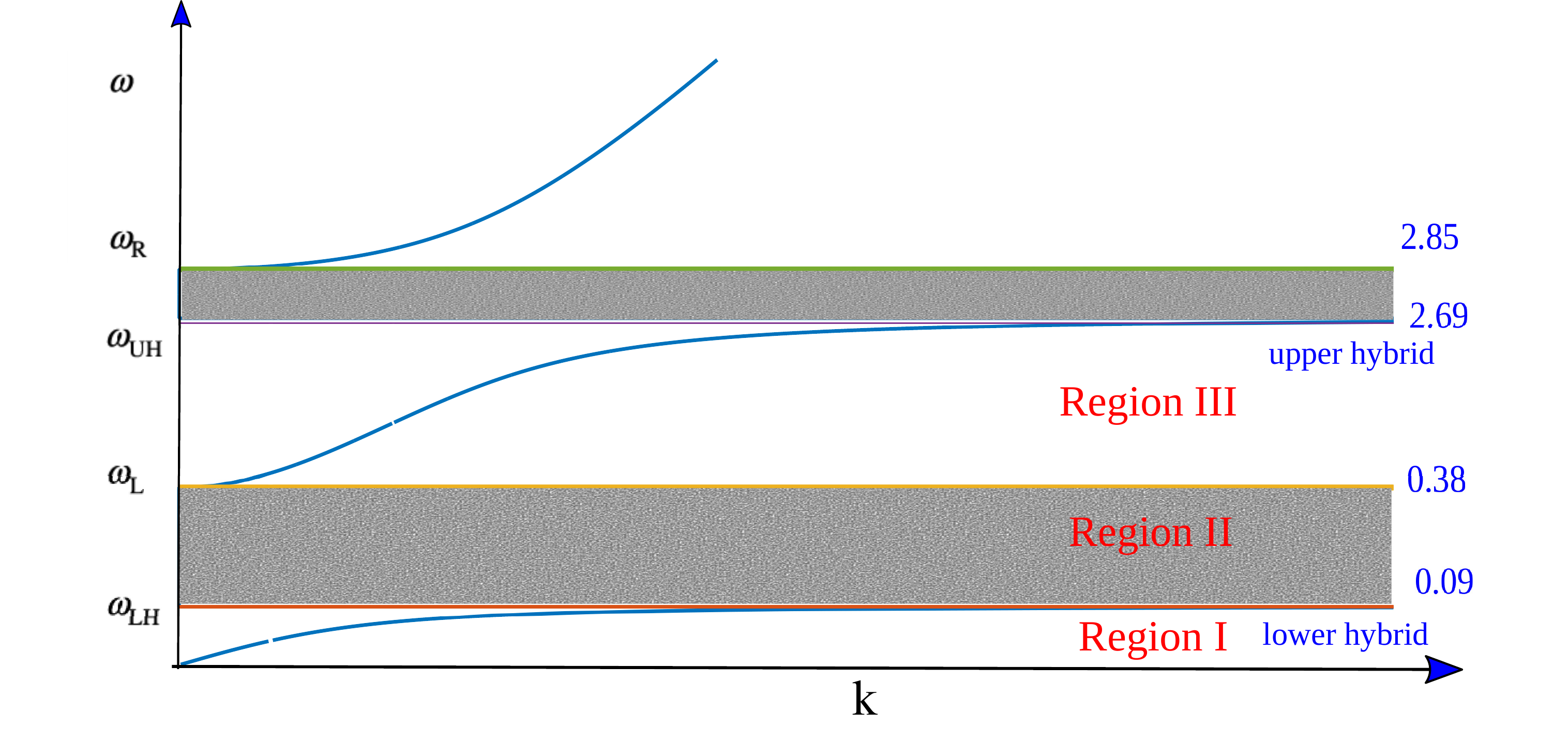}	

	\caption{Dispersion curve for X-mode propagation of EM wave in plasma. The curve shows different allowed and forbidden bands and their corresponding frequency limits. The numeric values in blue indicate the values of cut-offs and resonance frequencies for the plasma parameters considered in our simulation.}
		\label{dispersion}
\end{figure}

 \begin{figure*}
	\centering
	\textbf{\underline{(a) Variation of plasma density as well as lower hybrid resonance frequency with space}}\par\medskip
	\includegraphics[width=0.6\linewidth]{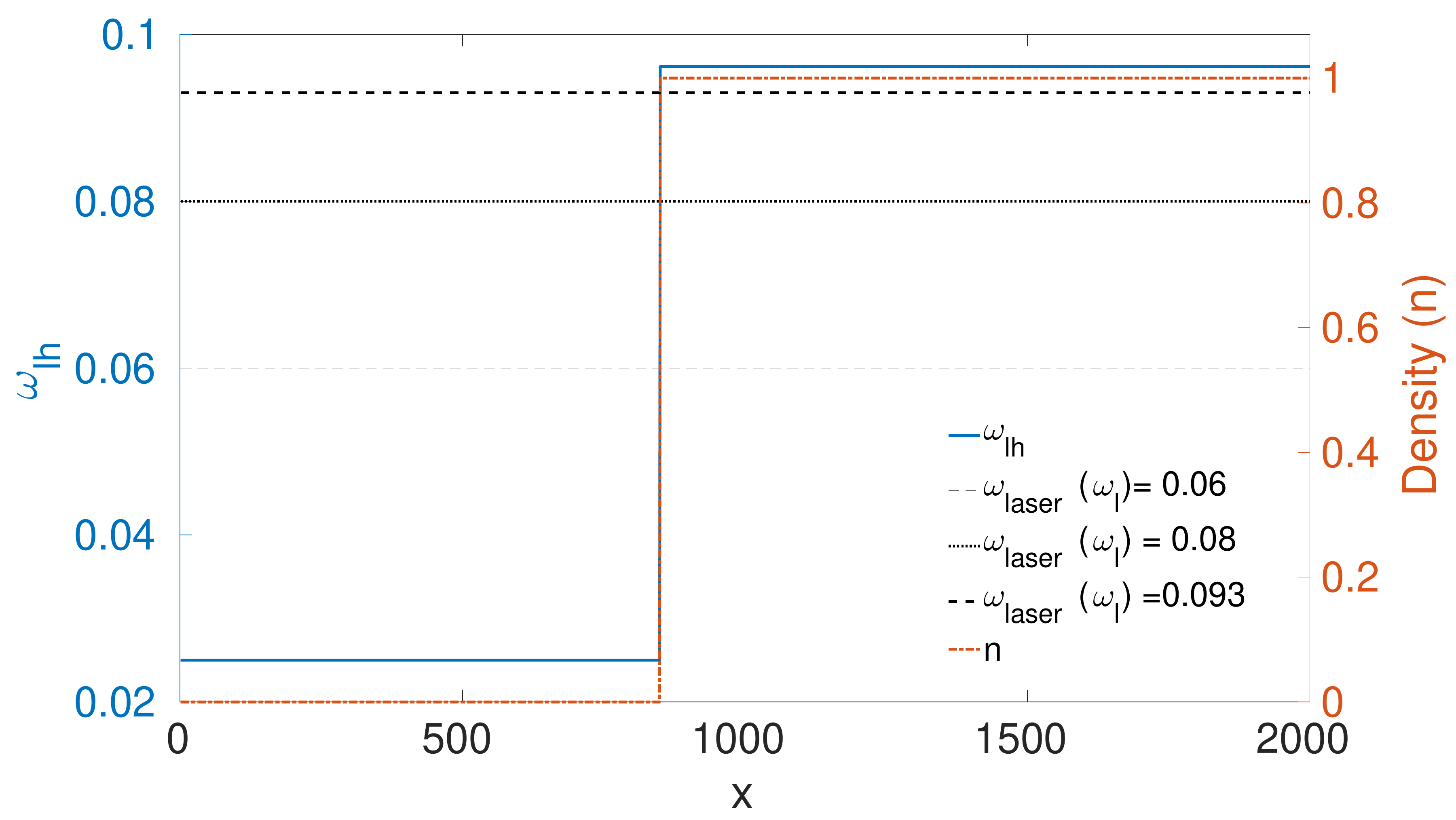}

		\textbf{\underline{(b) Electrostatic as well as magnetic field energy for different laser frequencies}}\par\medskip
		\includegraphics[width=0.7\linewidth]{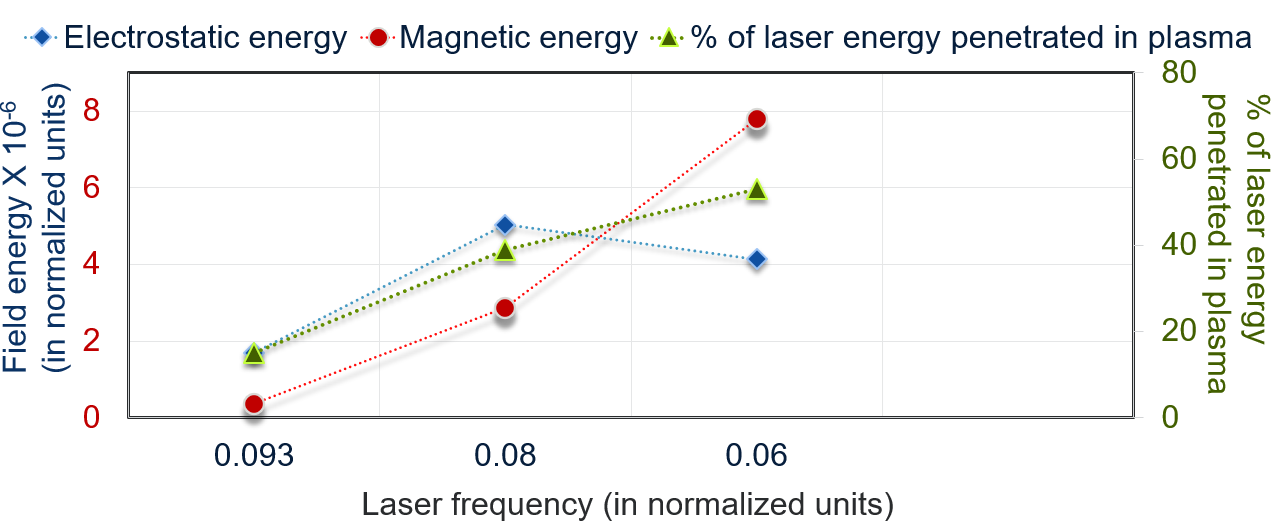}

	\caption{(a) The plasma density profile along with the variation of $\omega_{lh}$ with plasma density. (b) Field energies  and $\%$ of laser energy propagating inside plasma for three  laser frequencies i.e. $\omega_{l}=0.093$, $\omega_{l}=0.08$ and $\omega_{l}=0.06$. }
		\label{flat_3comp}
\end{figure*}

In our study, we have concentrated on X-mode configuration of electromagnetic wave interaction with plasma. For this configuration, we have resonances at lower hybrid and upper hybrid frequencies and cut-offs at left-hand and right-hand cut off frequencies (fig \ref{dispersion}). The expressions for the lower hybrid and upper hybrid resonance are given below \cite{boyd_sanderson_book}:

 \begin{equation}
\label{wlh}
\omega_{LH}= \sqrt{\frac{\omega_{ce} \omega_{ci} (\omega_{pe}^{2}+ \omega_{ce} \omega_{ci})}{\omega_{pe}^2 + \omega_{ce}^2 + \omega_{ci}^2}}
\end{equation}	
\begin{equation}
\label{wuh}
\omega_{UH}= \sqrt{\omega_{pe}^2 + \omega_{ce}^2 + \omega_{ci}^2}
\end{equation}

 where, $\omega_{ce} = |{eB/m_e e}|$ and  $\omega_{ci} = |{eB/m_i e}|$. Under the limit $\omega_{pe}> \omega_{ce}> \omega_{ci} $, the expression for $\omega_{LH}$ reduces to $\sqrt{\omega_{ce} \omega_{ci}}$.
The frequencies $\omega_L$ and $\omega_R$ shown in fig. \ref{dispersion} corresponds to the left and right hand cut off defined by the following expressions:
	\begin{equation}
\label{wl}
\omega_L= [\omega_{pe}^2+\omega_{pi}^2+(\omega_{ci}+\omega_{ce})^2/4]^{1/2} +(\omega_{ci}-\omega_{ce})/2
\end{equation}	
\begin{equation}
\label{wr}
\omega_R= [\omega_{pe}^2+\omega_{pi}^2+(\omega_{ci}+\omega_{ce})^2/4]^{1/2} -(\omega_{ci}-\omega_{ce})/2
\end{equation}

 Any incident frequency lying  between the resonance and cut-off value cannot penetrate inside plasma. Therefore,  there are  three pass and two stop bands for the incoming electromagnetic wave in X-mode dispersion curve. We have highlighted three of these bands marked with different regions (Fig \ref{dispersion}). In the work presented here, our focus is on laser frequencies lying below $\omega_{lh}$ i.e we want the laser to be lying in region I of the dispersion curve (fig. \ref{dispersion}). The choice of laser frequency in region I of fig. \ref{dispersion} is motivated from the fact that under the condition $\omega_{ce}> \omega_{l}> \omega_{ci}$, ions are able to respond and hence, gain energy from the incident laser \cite{vashistha2020new}.

We have considered two different configurations  (fig. \ref{density_profile3})  (I) a constant density profile, there being no  gradient. The plasma density is constant at $n=1$ in this case.   (II) a negative plasma density gradient is added to the plasma profile beyond the constant one i.e. plasma density starts from $n=1$, and then after some $\hat{x}$ starts falling linearly to $n=0.1$.  Falling plasma density profile has $n=1$ to $n=0.1$, dropping linearly with $\hat{x}$.  The discussion for both the profiles is organized in different sections.

\subsection{Propagation of laser as an electromagnetic mode in plasma for frequency lying below LH resonance;   A case study on Profile I}

 It has already been reported that when laser frequency lies close to the lower hybrid frequency, excitation of electrostatic lower hybrid mode takes place \cite{vashistha2021excitation}.  Further, to understand the propagation of laser below the resonance frequency, we consider three different laser frequencies incident on plasma.  One of the chosen laser frequency matches with the lower hybrid resonance frequency of the plasma ($\omega_{lh}= 0.093$) and the other two ($\omega_{lh}= 0.08$ and $0.06$) lie below it, in region I of the dispersion curve. The value of lower hybrid resonance frequency for the chosen simulation parameters is $\omega_{lh}= 0.093$. We plot saturated  field energies (at time $t= 2500 \omega_{wpe}^{-1}$), averaged over all space, for the given laser frequencies. We plot magnitude of $E_x^2/2$ as electrostatic energy and $B_z^2/2$ as magnetic energy in the system. The field energies are normalized to $m c \omega_{pe} e^{-1}$. It is to be noted that the three incident laser frequencies lie in region I of the dispersion curve, maintaining the criteria $\omega_{ce}> \omega_{l}> \omega_{ci}$.

\begin{figure*}
	\centering
\textbf{\underline{(a) Variation of plasma density as well as lower hybrid resonance frequency with space for Profile II}}\par\medskip
	\includegraphics[width=0.59\linewidth]{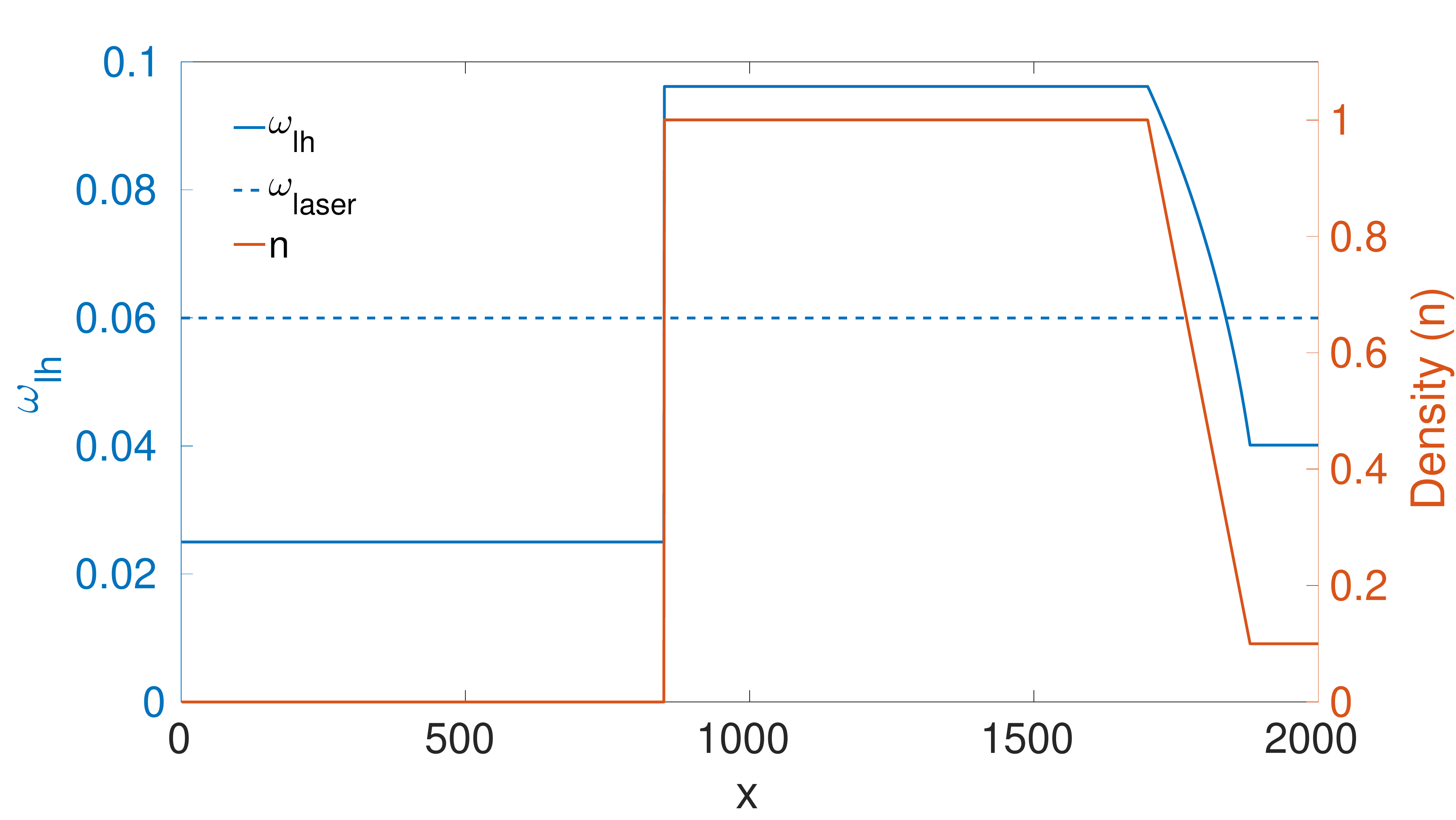}	
	
	\textbf{\underline{(b) Kinetic energy of ions in bulk plasma at three different times}}\par\medskip
	\includegraphics[width=0.59\linewidth]{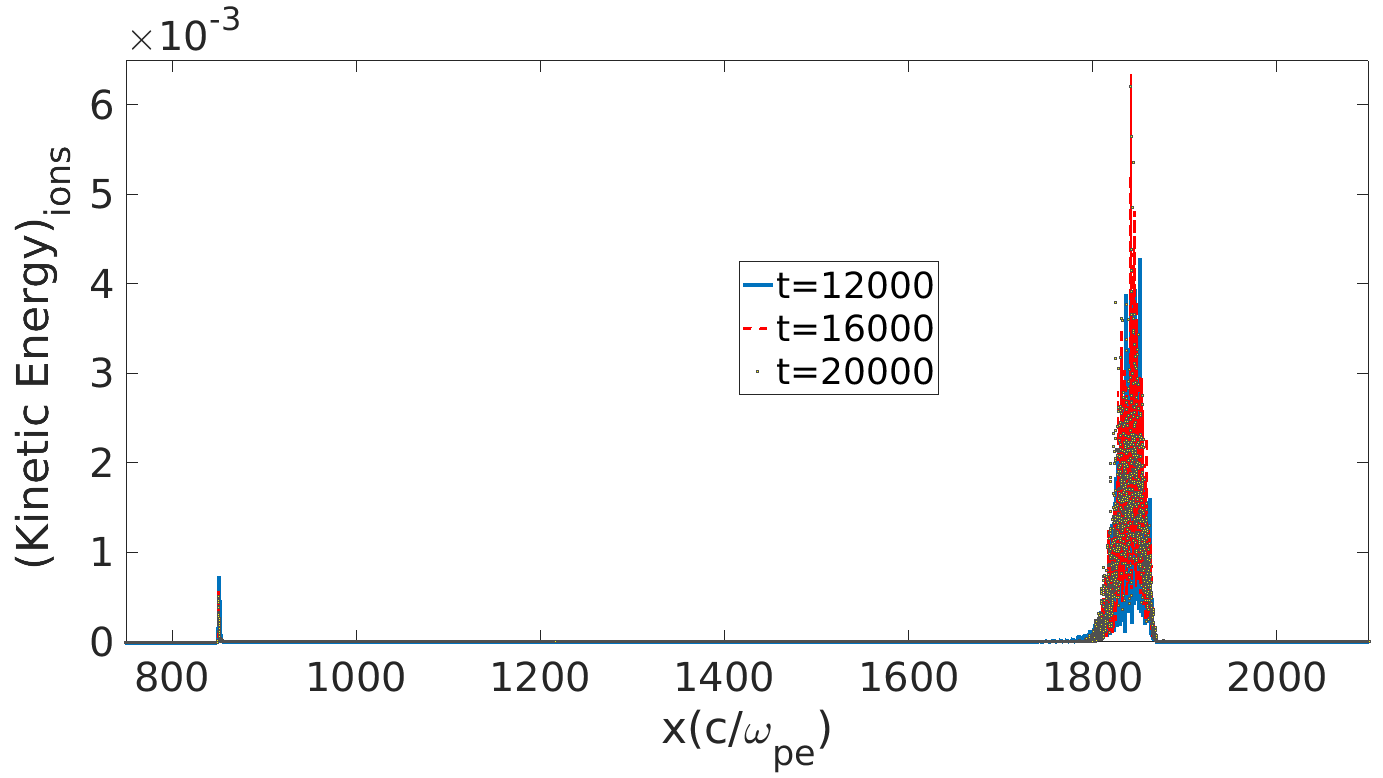}
	\caption{(a)Spatial variation of number density and corresponding lower hybrid frequency is shown. (b) Kinetic energy of ions in plasma at three different time is shown. The peak in kinetic energy is at the same location suggesting absorption of EM wave energy at a particular location in plasma. }
		\label{profile_ke_comp}
\end{figure*}

 Fig. \ref{flat_3comp} (b) lays out two major inferences for simulation runs in region I of dispersion curve. (i) Penetration of laser in plasma increases with decreasing the laser frequency. We observe that more laser is able to penetrate and hence propagate inside plasma if its frequency lies farther below the lower hybrid resonance value (as shown by green markers). (ii) The electrostatic component of field energy (magnitude of $E_x^2/2$) in plasma dominates when laser frequency is close to lower hybrid frequency ($\omega_{l} = 0.093$) and decreases as it goes farther below (shown by blue markers). The electrostatic component in plasma dominates over the electromagnetic one for $\omega_{l}=0.093$. However, the situation reverses for lower incident frequency ($\omega_{l}=0.06$). Upon decreasing the laser frequency, magnetic energy surpasses the electrostatic energy. Therefore, one can conclude that upon moving farther below the resonance frequency, the laser propagates as an electromagnetic wave in plasma. This electromagnetic wave will continue to propagate inside plasma, without any irreversible energy transfer into plasma species. Hence, to harness energy from this electromagnetic wave, we introduce a gradient in plasma density. The gradient is chosen in such a way that the resonance point for the EM wave lies inside the bulk plasma. The details  are discussed in the section to follow.

\begin{figure*}
	\centering
	\textbf{\underline{Electromagnetic field energy with space and time}}\par\medskip
	\includegraphics[width=0.79\linewidth]{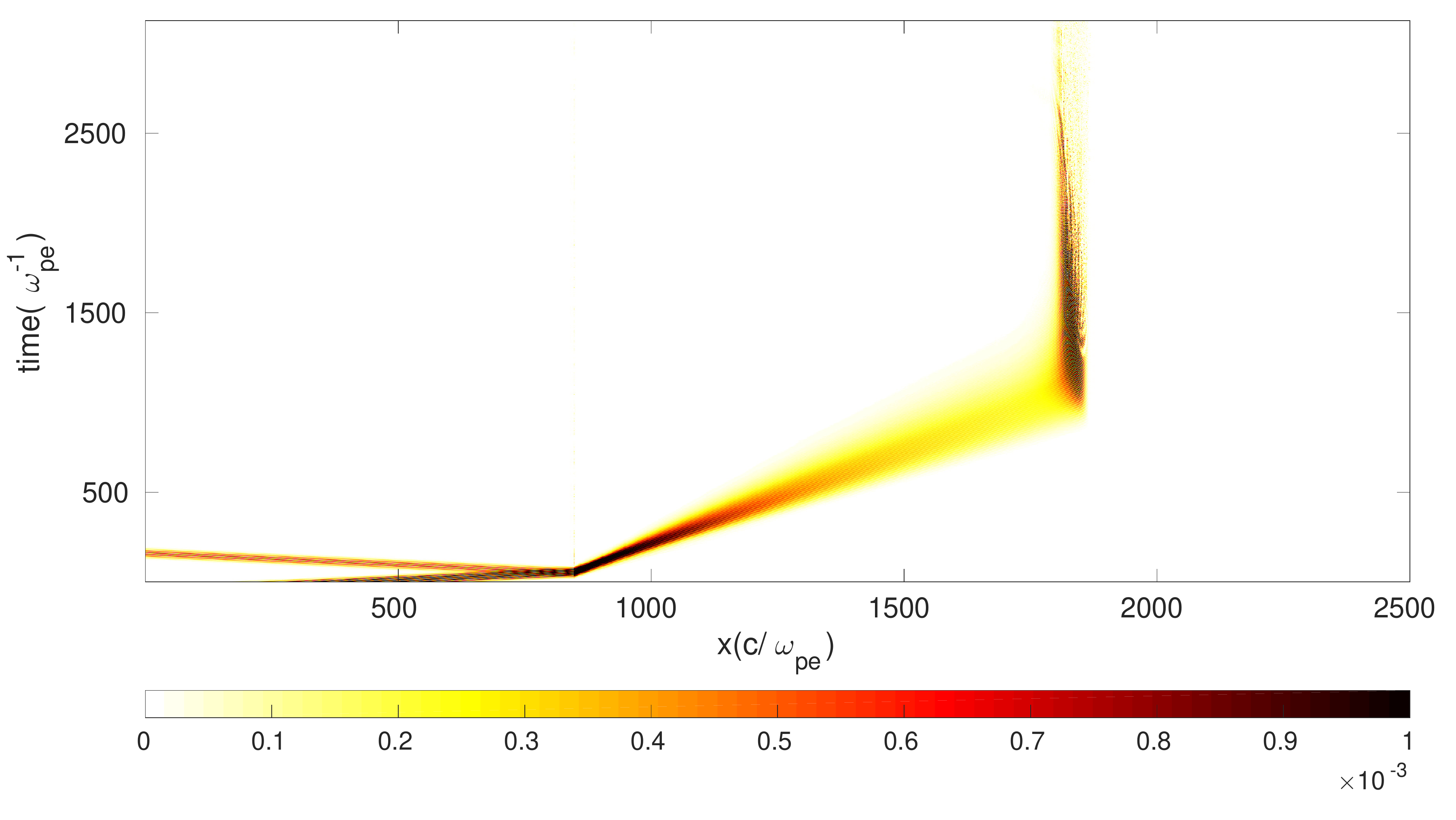}	
		\textbf{\underline{Kinetic energy of ions with space and time}}\par\medskip
\includegraphics[width=0.79\linewidth]{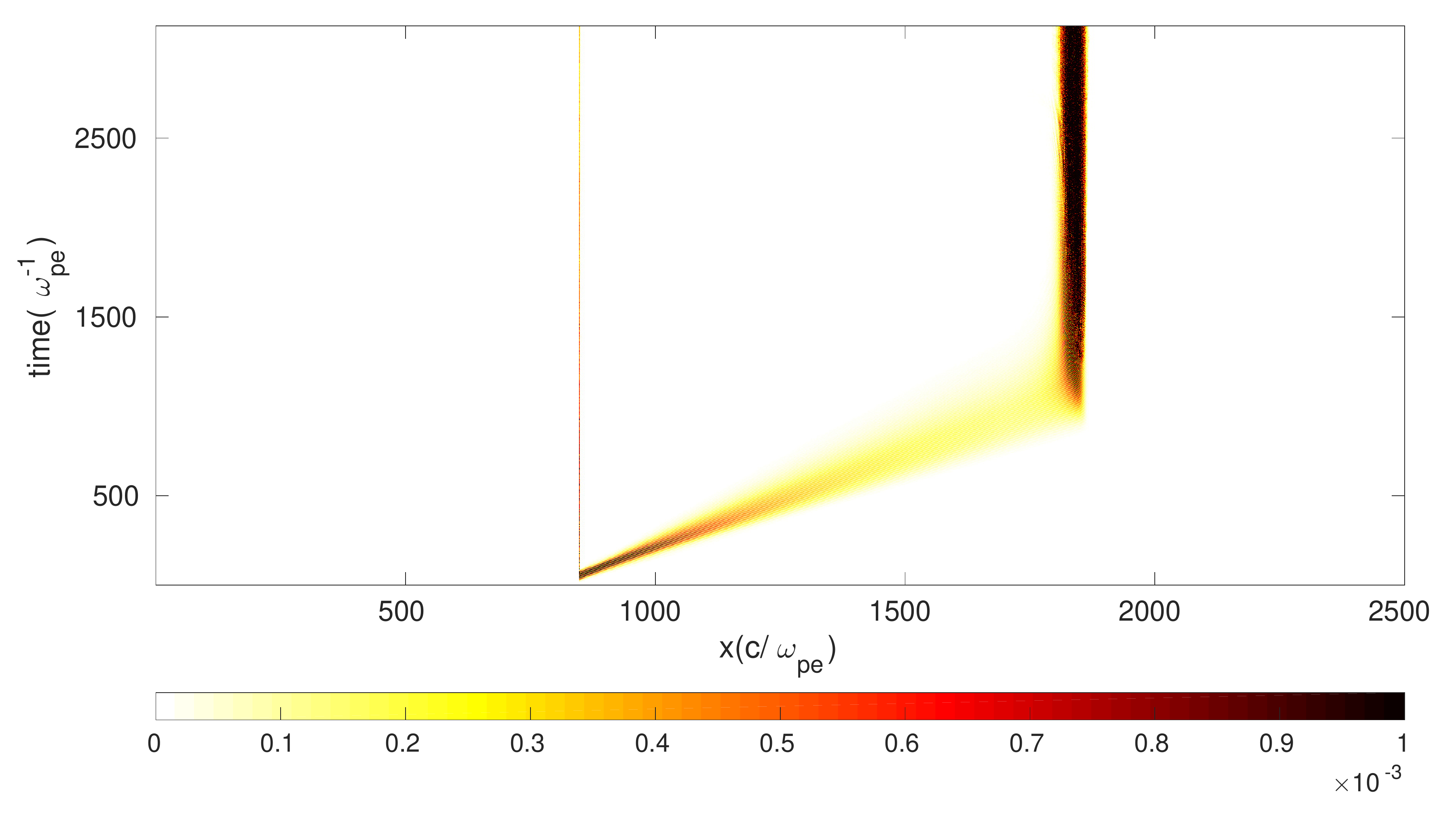}	
	\caption{Variation of field energy and kinetic energy of ions with space and time. We observe a part of laser propagates inside plasma until it encounters the resonance point. Similarly, we observe a finite energy in ion species in plasma with the propagation of laser. However as the resonance point is reached, we observe a drop in field energy and a rise in kinetic energy of ions. }
		\label{energy_x_t}
\end{figure*}

\begin{figure*}
	\centering

\includegraphics[width=0.79\linewidth]{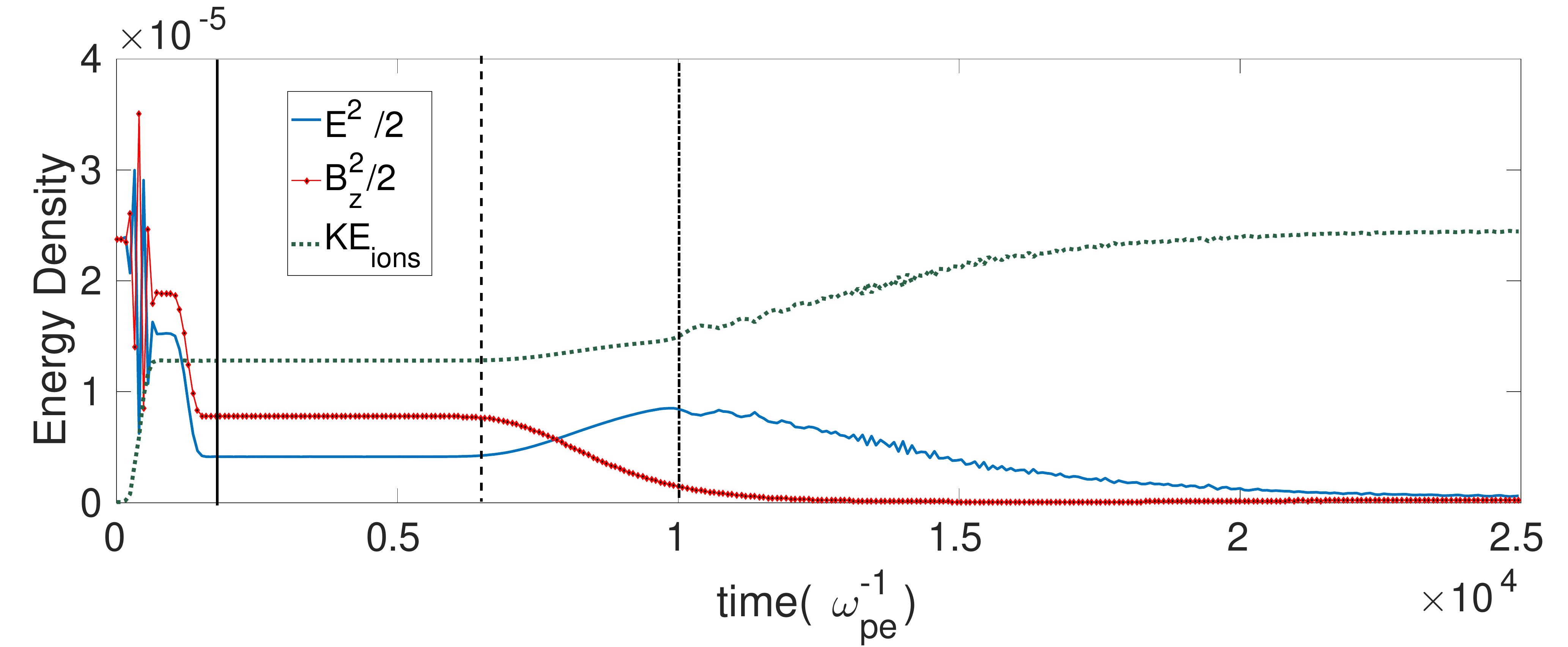}	
	\caption{Variation of field energies and Kinetic energy of ions with time for a constant and then falling plasma density profile. Laser interacts with plasma and a part of it is propagating inside as an electromagnetic mode.  The field and kinetic energies get saturated between the solid and dashed line as it propagates in the constant plasma density profile. As these oscillations encounter a falling plasma density, we observe a fall in $B_z^2$ and a simultaneous rise in $E^2$ (between dashed and dotted line). This suggests that electromagnetic mode is now converting to electrostatic mode. Further we observe conversion of electrostatic mode into kinetic energy of ions (beyond dotted line). This suggest the presence of a resonance point in plasma where conversion of electromagnetic to electrostatic and further into kinetic energy into ions takes place. }
		\label{ene_ff_wlpo6}
\end{figure*}

\begin{figure*}
	\centering

\includegraphics[width=0.79\linewidth]{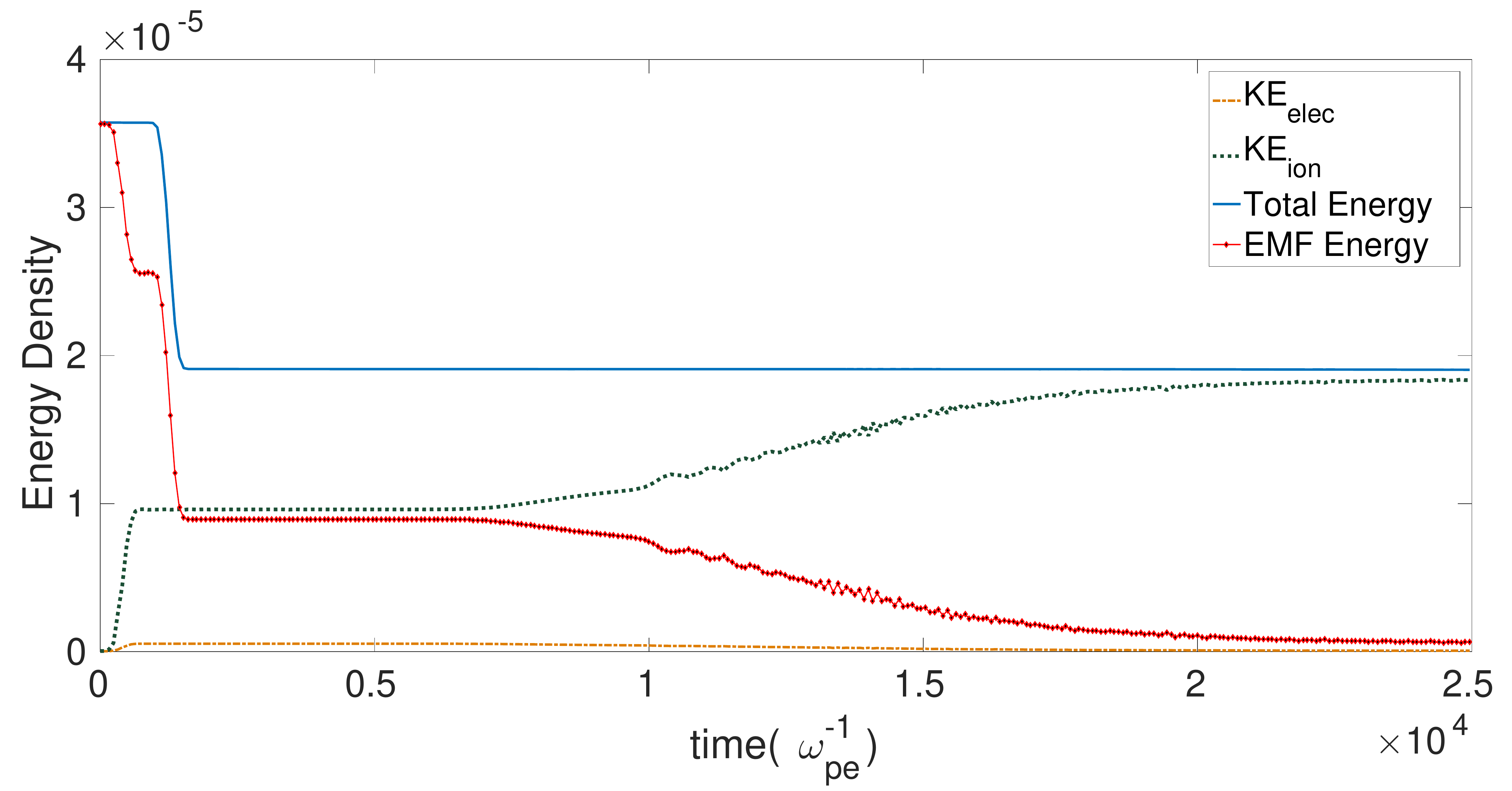}	
	\caption{Variation of field energies and kinetic energy of electrons and ions with time for Profile II. EMF energy is the total of electric and magnetic field energies.  The plot shows that the total energy is conserved in the system. Therefore,  the results presented in our simulation are physical. }
		\label{tot_energy_ff200}
\end{figure*}

\subsection{Localized absorption of laser energy  at lower hybrid resonance point in plasma; A study on Profile II}

In this section, we study the effect of negative density gradient in conversion of electromagnetic energy from laser into kinetic energy of plasma species, particularly ions. We introduce a negative gradient in plasma density as shown in fig.\ref{profile_ke_comp}(a). The laser frequency is chosen to be $\omega_l = 0.06$ (in normalised units). Fig. \ref{profile_ke_comp} (b), indicates presence of a resonance point in plasma. We plot  kinetic energy of ions as a function of space for three different times and  observe peak in kinetic energy at a particular location in plasma at all times. This suggests that all the electromagnetic energy from laser has been converted to kinetic energy of ions at this particular location in plasma. 

To further investigate, we plot variation of electromagnetic energy (fig. \ref{energy_x_t}(a)) and kinetic energy of ions  (fig. \ref{energy_x_t}(b)) as a function of space and time. We observe a decrease in field energy and a simultaneous increase in ion energy at a particular location in plasma. 
This is indicative of an energy transfer from laser to plasma species at a designated location. This location is the lower hybrid resonance point for the given plasma parameters. As the laser is incident on plasma, it first encounters a constant plasma density. The part of the laser which has entered the plasma, propagates inside. Thereafter, this propagating mode encounters a fall in the plasma density. This is where the EM mode finds a resonance point at the lower hybrid frequency of plasma. The natural frequencies of plasma (e.g $\omega_{pe}$, $\omega_{pi}$, $\omega_{lh}$) all vary with plasma density. Hence, there lies a point on the density profile where the propagating wave frequency matches or lies close to the lower hybrid frequency. The wave cannot propagate beyond this point as there are no propagating solution between $\omega_{lh}$ and left-hand-cutoff frequency ($\omega_{L}$) (can be observed from the linear dispersion curve in fig. \ref{dispersion}). The dispersion curve (fig. \ref{dispersion}) at the resonance point shows a saturation with lower hybrid mode being the solution of the curve. Hence, the wave stops as soon as it reaches the resonance point. At the lower hybrid resonance point, all the electromagnetic energy gets converted to electrostatic. Further, the conversion of electrostatic energy into kinetic energy of ions takes place. The conversion of electrostatic energy into plasma species is due to collisionless interaction of wave with the particles at that location. In our system, electrons  are tightly bound to the external magnetic field, as a result, ions take energy from the wave at the resonance location.


To understand the role-play of various energy components in transfer of field energy, we plot electric field energy ($E^2/2$), magnetic energy ($B_z^2/2$) and  ion energy, averaged over space, as a function of time (fig. \ref{ene_ff_wlpo6}).  Initially we observe propagation of an EM mode in plasma (between solid and dashed line).  The EM mode then encounters a falling plasma density. There comes a point when the frequency of the propagating electromagnetic mode approaches the lower hybrid resonance point. This is where the conversion of electromagnetic to electrostatic energy takes place (between dashed and dashed dotted line). Further in time, we observe a decrease in $E^{2}/2$ as well and an increase in kinetic energy of ions. Thereby, we observe maximum transfer of energy from laser to plasma species through a two stage process, $(i)$ conversion of electromagnetic energy into electrostatic energy. $(ii)$ breaking of electrostatic oscillations to impart energy to ion species.  Further, we plot the total energy of the system to show that the results presented are physical (Fig. \ref{tot_energy_ff200}). Fig. \ref{tot_energy_ff200} also shows that laser energy is dominantly dumped into ions species, whereas electrons being tied to the magnetic field (under the condition $\omega_{ce}> \omega_{l}> \omega_{ci}$) do not respond the the incoming laser.

\subsection{Effect of density gradient on energy conversion efficiency}

\begin{figure*}


\includegraphics[width=0.59\linewidth]{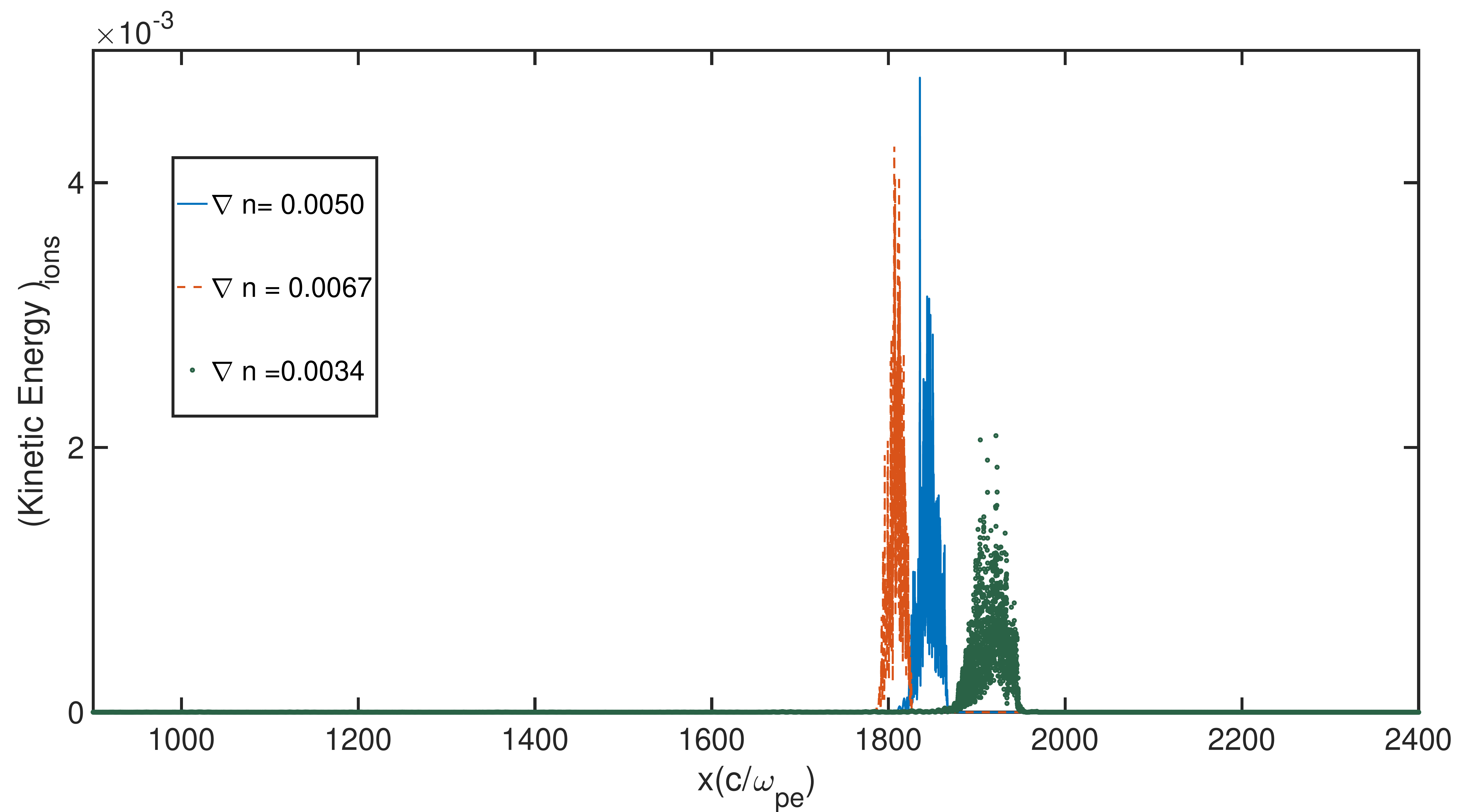}	
	\caption{Kinetic energy of ions for different plasma density gradients. One can observe a shift in the peak location for the three different gradients. This is because of the shift in the resonance point with plasma density. }
		\label{peak_shift}
\end{figure*}

To understand the effect of energy conversion on plasma density gradient, we consider three different density gradients ($\nabla n = 0.0034, 0.005, 0.0067$). We observe that the conversion efficiency for all the three gradients is the same. This will however change if we change the density gradient to too sharp or too shallow gradients.  A comprehensive study on gradient scale length for upper hybrid resonance absorption by Maity et. al. \cite{maity2022mode} show that conversion efficiency to electron kinetic energy, in their case decreases when magnetic field gradient is steep. This is because  they observe increase in efficiency of other two mechanisms, which are, parametric decay instability  and generation of higher harmonics at the UH resonance layer. This leads to conversion of electromagnetic energy into these two mechanisms and hence conversion efficiency due to upper hybrid reduces. The gradient scale length chosen in our simulations on the other hand are in the range where ion response does not change significantly. Hence, we observe no change in conversion efficiency of laser into ion energy  by changing plasma density gradient. This will however change if the gradient scale length become comparable to ion response time scales. Further, we plot kinetic energy of ions for all these cases, as a function of space. Fig. \ref{peak_shift}  is a snapshot of the time when the energies in the system have saturated. We observe that the peak point for ion energy (Fig. \ref{peak_shift}) shifts.  This is expected because the resonance point will shift with the gradient in plasma density.

\section{Conclusion}

We have studied the propagation of an incident EM pulse in X-mode configuration on  different plasma density profiles. We demonstrate the propagation of laser in plasma when its frequency lies below the lower hybrid resonance value. Further, we show the conversion of electromagnetic energy into kinetic energy of ions using suitable density gradients in plasma. This work proposes a mechanism for absorption of laser energy into ions species at a designated location in plasma. This can have application where localized absorption of laser energy is required. Furthermore, we contribute to the fundamental understanding of laser propagation in inhomogeneous plasma in the presence of  an externally applied magnetic field. With the recent advancements in achievement of kilo-Tesla magnetic field \cite{Nakamura} and  low frequency, short-pulse, $CO_2$ laser, this study of laser with magnetized plasma should be well within reach of experimental campaigns.\\



 \section*{Acknowledgements}
 The authors would like to acknowledge the OSIRIS Consortium, consisting of UCLA and IST(Lisbon, Portugal) for providing access to the OSIRIS4.0 framework which is the work supported by NSF ACI-1339893. AD
would like to acknowledge her J. C. Bose fellowship grant JCB/2017/000055 and the CRG/2018/000624 grant of DST for the work. The simulations for the work described in this paper were performed on Antya, an IPR Linux cluster.

\paragraph*{\bf{{References:}}}

\bibliography{LH_reso}

\end{document}